\documentstyle[12pt]{article}
\begin{document}
\titlepage
\title{\begin{flushright}
Preprint LNPI-1099,  1985
\end{flushright}
\bigskip\bigskip\bigskip
ELECTROMAGNETIC FIELDS PROVIDING CHARGE SCREENING AND
CONFINEMENT IN TWO-DIMENSIONAL MASSLESS
ELECTRODYNAMICS\thanks{The brief version of this paper has been
published in Yadernaja Fizika: Yad. Fiz. 41 (1985) 534 (in
Russian).}} \date{}
\author{G.S.  Danilov \thanks{E-mail address:
danilov@thd.pnpi.spb.ru}, I.T. Dyatlov\thanks{E-mail address:
dyatlov@thd.pnpi.spb.ru} and V.Yu.Petrov\thanks{E-mail address:
victorp@thd.pnpi.spb.ru}\\ Petersburg Nuclear
Physics Institute,\\ Gatchina, 188350, St.-Petersburg, Russia}

\maketitle
\begin{abstract}
The charge screening, confinement of fermion quantum numbers
and the chiral condensate  formation in two-\-dimensional QED is
studied in details.  It is shown that charge screening and
confinement of fermion number in two-\-dimensional QED is due to
an appearance of gauge fields which nullify the Dirac
determinant $D(A)$.  An appearance of the fields of another type
but with the same property yield the chiral condensate
formation. In addition, these second type fields ensure the
"softness" of the charge screening in a process  which is
analogous to the $e^+e^-$ annihilation.
\end{abstract}

\newpage

\section{Introduction}

The conventional point of view [1] connects quark confinement
with a linear potential which is expected for two static quarks
inserted in the vacuum of pure gluodynamics. It is usually
assumed that the colour field of the quarks interacting with
non-\-perturbative gluon vacuum fluctuations shrinks into the
tube. This leads to a linear growth of the energy with the
distance between quarks and confines them. The situation could
drastically change, however, when light quarks are introduced in
the theory. Here the string between static quarks instantly
tears and created quark pairs screen the linear potential of the
pure gluodynamics. This means that the specific properties of
QCD with light  quarks are determined by configurations of gluon
fields rather different from the configurations providing
confinement in pure gluodynamics.

The simple example of the phenomenon discussed is the
well-\-known Schwinger model presenting the quantum
electrodynamics of massless fermions (quarks) in two dimensions
[2,3]. Here the original Coulomb potential is indeed
proportional to the distance but the interaction with massless
quarks screens it out. As the result, one can observe in
QED$_2$
\begin{description}
\item[i)] the absence of
charged (coloured) states [2--4];
\item[ii)]  confinement
of quark quantum numbers (chirality, fermion number [5], flavour
for multiflavour QED$_2$ [4,6]);
\item[iii)] the chiral
vacuum condensate [3,6,7];
\item[iv)] the independence of
quark jets and "soft" hadronization [8,9].
\end{description}
The above features of
QED$_2$ seem to be similar to that expected in QCD.

In this paper we just investigate the electromagnetic
(e.m.) fields which provide the above features of QED$_2$. We
show that these are fields for which the Dirac
operator determinant $D(A)$ vanishes. We divide the above
fields in two types to be, respectively, A and B.

The fields of the first (A) type are necessary generated in
QED$_2$ by charged systems. They are responsible for charge
confinement in QED$_2$. In order to prove this we consider the
evolution operator $S(T)$ of the system in the quark
representation.  This operator is the sum over all physical
states
\begin{equation}
S(T)=\sum_h e^{-iE_nT}|\Psi_n><\Psi_n|,
\end{equation}
$E_n$ being the energy of the state. Matrix elements of the
operator $S(T)$, Eq.(1), can be represented as functional
integrals over the fields with boundary values fixed at initial
and final momenta of time. In the Coulomb gauge which we shall
use in the paper we have:\footnote{This gauge in convenient
since here are no unphysical degrees of freedom. This gives
possibility to present a clear physical interpretation of all
phenomena considered}
\begin{equation}
S(T)=\int
DA_0\exp\left[-i\int\limits^T_0\frac{F^2_{\mu\nu}}4
d^2x\right]\int\limits^{\Psi_{out}(T)}_{\Psi_{in}(0)}D\Psi
D\bar\Psi \exp\left[i\int^T_0\bar\Psi(i\hat\partial+g\hat
A_0)\Psi d^2x\right].
\end{equation}
For QED$_2$ we have constructed
$S(T)$, in ref.[10].

Matrix elements of the evolution operator determine the
probability amplitudes to have any quark configuration in any
physical state. The charge confinement means that such
amplitudes must vanish for the configurations where some charge
(quark $q$) is situated far from other compensating charges
(antiquark $\bar q$). We shall demonstrate that this is the case
in QED$_2$ and the amplitude to find quark on distance $r$ from
antiquark falls as $r^{-2}$. Investigating the structure of the
integral, Eq.(2), one can discover the reason for such a
dependence: the self-\-consistent field $\bar A$ of the
distantly situated quark annihilation the quark determinant $D(A)$.
This is the field of the class A. Since the amplitude is
proportional to $D(\bar A)$ it naturally vanishes at large $r$.
The explicit proportionality to $D(A)$ appears in Eq.(2), after
integration over fermion fields. We have [10]:
\begin{equation}
S(T)=\int
DA_0\exp\left[-i\int^T_0\frac{F^2_{\mu\nu}}4\,d^2x\right]
S_{ext}(A_0).
\end{equation}
Here $S_{ext}(A_0,t)\sim D(A_0)$ is the evolution operator in the
given external field $A_0$ (explicit form see in Sec.2). Its
matrix elements are determined by chosen quark configurations
and are expressed in terms of the fermion Green functions in the
external e.m.  field $A_0$ (Eq.(5)).

Now we see that if for some quark configuration a self-consistent
field $\bar A_0$ nullifies $D(A)$ then this configuration is
prohibited; it cannot appear in evolution of the system. It is
this mechanism which forbids existence of charged states in
QED$_2$. As we have stated the field of every charged quark
creates the field ,which makes determinant to vanish.  The field
strength of the discussed field behaves as $1/x$ for large
distances due to the effect of the vacuum polarization.

The fields with $D(A)\to0$ always create new quark pairs.
\footnote{In fact, $|D(A)|^2$ is the probability of the process
where no quark pairs are produced.} These pairs screen original
charges. If the total screening is not achieved, a new
configuration also corresponds to $D(\bar A)\to0$ and the
process repeats. It is well known that $D(A)$ determines vacuum
polarization effects. Hence, the described processes represent
the mechanism of charge screening by means of the vacuum
polarization.

As it had been noted by many authors [6,11,12], the screening
mechanism of QED$_2$ reminds in some aspects the well-known Higgs
one. However, there is an essential difference between them. The
QED$_2$ models contain no charged condensate because any
configuration with separated charges cannot be present in the
evolution operator $S(T)$. Therefore, there are no charged
complexes in the physical vacuum. We had seen this fact
investigating the quark structure of QED$_2$ vacuum states [7].
Charge screening is not statistical (in average) as it happens
in the Higgs case, but is the exact one. The wave functions of
excitations are here the eigenfunctions of the charge operator
with vanishing eigenvalues. It is just the spectrum of the type
expected for the real world [1].

The fields of the second class (B) govern chiral properties of
the model. They also make $D(A)\to0$, but decrease with a space
distance faster than $x^{-1}$. They can be characterized by a
non-\-zero value of the integral
\begin{equation}
Q_{top}(T)=-\frac
g{2\pi}\int\limits^T_0\int d^2x\,E(x,t)=-\frac g{4\pi}
\int\limits^T_0\int d^2x\varepsilon_{\mu\nu} F^{\mu\nu}(x),
\end{equation}
which is the two-dimensional winding number [13]. The $B$-fields
also inevitably produce $q\bar q$-\-pairs. As we
shall see $2|Q_{top}|$ particles of these pairs necessarily have
very small momenta $p\sim1/V$, being considered in the finite
space volume $V$ ($V\to\infty)$. They are delocalized over the
whole volume. Such particles do not contribute to any local
quantities and are, in fact, unobservable. In the $V\to\infty$
limit, the non-\-conservation of chirality $K$ (which follows
from the Adler-\-Bell-\-Jackiw (ABJ)-\-anomaly [14]) can be
interpreted in QED$_2$  just as a generation of delocalized
particles with small momenta $p\sim1/V$ by $B$-fields.

The $B$-fields with integer $Q_{top}$ play an important role in
the formation of quark condensates. For $Q_{top}=N$ exactly $N$
pairs $\bar q_Rq_L$ (or $q_R\bar q_L$ for $N<0$) are generated
by such a $B$-type field. The chiral condensates of QED$_2$ with
one and several flavours [2,6] are formed by similar chiral
complexes [7] (Sec.3).

We shall see in Sec.3 that $B$-fields prevent a change of
chirality for developing quark configurations. Owing to this
property the exchange of chirality between hadrons and the
chiral condensate is impossible. As a result, chirality of
hadrons is conserved and appears to be the definite fixed
quantity $(K=0)$.

In some aspects the properties of B-fields are similar to those
of instanton fields [15] in QCD. However there are essential
differences between them, as  it is discussed  in
Sec.2.

The second main topic of this article is the hadronization
process investigated in Sec.4. We consider there the
production of a $q_R\bar q_L$-\-pair by an external source.
This, widely discussed in QED$_2$, process [3,8,9] is an
analogue of the $e^+e^\rightarrow$hadrons reaction.
It is also governed by a field with $D(A)\to0$
(B-type). Such a field produces just that one $\bar q_R
q_L$-\-pair with finite momenta $(p\sim m)$ which screens
charges and chiralities of the both jets. Simultaneously, the
same field creates another $q_R\bar q_L$-\-pair with the
vanishing total momenta which
joins the vacuum condensate.

Thus the screening of charges and chiralities for both
jets is finished at the finite time $t\sim m^{-1}$ ($m=g^2/\pi$
is the hadron mass), when $Q_{top}(T)$ becomes of the order of
unity.  At larger times $t>1/m$ we have only neutral independent
jets with $K=0$. From these times, there begins  a
rapid growth (proportional to $t^2$) in the number of quarks
with momenta $p\gg m$. At times $t\sim p/m^2$ their distribution
is transformed into the hadron distribution of the parton model
$(1/p)$.  The leading pair of quarks loses its energy, and after
a time $t\sim p_{in}/m^2$ participates in the formation of the
hadrons.  The produced hadrons with $p< p_{in}$ become spatially
separated from the leading quarks of the initial pair. Quarks
exist outside hadrons only on the light cone.

\section{Properties of fields where $D(A)\to0$}

EM fields which  make the determinant $D(A)\to0$ correspond to
those quark configurations which are prohibited for evolution
and production. In order to prove this statement let us consider
the evolution operator $S(T)$ for a finite time interval $T$. As
it was shown in Ref.[10], the operator $S(T)$ can be represented
as a functional integral over $A$. In the Coulomb gauge
($A_1=0)$ we have Eq.(2) for $S(T)$.

The integration in Eq.(2) goes over all EM fields $A_0$ in the
time interval $0\le t\le T$ including end points $t=0$ and
$t=T$. The evolution operator for massless fermion in an
external field $A_0$ is equal to [10]
\begin{eqnarray}
S_{ext}(T)&=&D(A_0)\exp\left\{\int\limits
dx\,dx'\sum_{i=R,L}\left[
a^+_i(x)G^{(T)}_i(xT,x'0)a_i(x')\right.\right. \nonumber\\
&+& b_i(x)G^{(T)}_i(x,0;x',T)b^+_i(x')-a^+_i(x)G^{(T)}_i
(x,T-\varepsilon;x'_,T)b^+_i(x') \nonumber\\
&-&\left.\left. b_i(x) G^{(T)}_i(x,\varepsilon;x',0)a_i(x')\right]
\right\}; \\
\varepsilon & \to& +\;0. \nonumber
\end{eqnarray}
Here $D(A_0)$ is the determinant of the Dirac equation in the
external field $A_0$; $a^{\pm}_{RL}(x)$, $b^{\pm}_{RL}(x)$ $(a^-
\equiv a)$ are the creation and annihilation operator for right
$(R)$ and left $(L)$ quarks and antiquarks:
\begin{equation}
\Psi_i(x)=a_i(x)+b_i^+(x), \qquad i=R,L, \qquad
\Psi_{R,L}(x)=\frac12(1\pm\gamma_5)\Psi(x).
\end{equation}
The Green functions $G_i$ in an external field have the form [10]
\begin{eqnarray}
G_i^{(T)}(x,t;x',t')&=&G^{(0)}_i(x,t;x',t')\exp\left\{
ig\int\limits^T_0 dt_1\int dx_1 A_0(x,t)\right. \nonumber \\
& \times & \left.\frac{}{}\left[ G^{(0)}_i(x,t;x'_1,t_1)-
G^{(0)}_i(x',t'; x_1,t_1)\right]\right\}, \nonumber \\
G^{(0)}_i(x,t;x',t')&=
&[2\pi i(t-t'\mp (x-x)-i0\mbox{ sign }(t-t'))
]^{-1},
\end{eqnarray}
$g$ is the dimensional coupling constant of QED$_2$.

For any processes, matrix elements of $S(T)$ are determined by
quark configurations in the initial $(a_i(x),b_i(x))$ and final
$(a^+_i(x), b^+_i(x))$ states. These matrices appear in Eq.(3) as
coefficients in front of corresponding numbers of $a^{\pm}$,
$b^{\pm}$-operators. To obtain matrix elements we must expand
the exponential in Eq.(5). Then the integral, Eq.(3), will be of
the Gauss-\-type one. Its value can be obtained substituting the
saddle-\-point field $\bar A_0$, in every term of the
integrand.

Hence, if $D(\bar A_0)\to0$, the configuration discussed does not
contribute in the norm of any physical state vector:
\begin{equation}
\Psi(T)\; =\; S(T)\Psi_0.
\end{equation}
It does mean that this configuration cannot appear as a result of
evolution of the system. If charged states exist, then distantly
situated charges would be represented in matrix elements of the
$S(T)$-\-operator with a finite probability. There are no charged
states when all such matrix elements decrease with space
distance $r$ between charges. As we shall see, in QED$_2$ all
the above fenomena are due to fields providing $D(A)\to0,$ at
$r\to\infty$.

In this section we discuss general properties of the fields
$D(A)\to0$. We take here fields as external ones.
Self-\-consistent fields of QED$_2$ considered in the next
section, are of the $D(A)\to0$ type, too.

The Dirac determinant for massless fermions in an external field
has the following form [10]:
\begin{equation}
\ln D(A)=-\frac{m^2}2\int\frac{dp}{2\pi}\int\limits^T_0 d\,td\,t'
E(p,t)E(-p,t)\frac{e^{-i|p|\,|t-t'|}}{2|p|}.
\end{equation}
$m^2=g^2/\pi$, $E(x,t)=-\partial A_0/\partial x$ is the field
strength in the Coulomb gauge.

All the field configurations discussed correspond to
finite field strength. Then $D(A)$ can vanish only because of the
divergence of the integral in Eq.(9) at $p\to0$. Therefore, we
shall consider two types of the field $D(A)\to0$
\begin{equation}
{\rm (A) }\; E(p,t)\to{\rm sign } p\,f_1(t), \qquad {\rm (B) } \;
E(p,t)\to f_2(t),\; p\to0,
\end{equation}
where $f_i(t)$ are some functions of $t$. These fields have
different asymptotics for $x\to\infty$. The field strength for
the A-type fields decreases as $1/x$, while $E$ for the B-type
ones decreases more rapidly.

Evaluating a divergent part of the integral, Eq.(9), we obtain the
following formulae for the $D(A)$ using space volume $V$ as an
infrared regularization:
\begin{equation}
D(A)=1/V\beta^{2(T)}, \quad \beta(T)=\frac g{2\pi}\int^T_0 dt\int
dxV.P.\int \frac{dyE(y,t)}{x-y}
\end{equation}
for the A-type fields and
\begin{equation}
D(A)=1/V^{|Q_{top}(T)|}, \quad Q_{top}(T)=-\frac
g{2\pi}\int^T_0dt\int dx\,E(x,t)
\end{equation}
for the B-type. $Q_{top}$ is the "topological charge", of the field
(see Sec.1).

The main contribution to the norm of state vectors for the fields
$D(A)\to0$ comes from the particles with small momentum $p\sim1/V$.
Indeed, since the norm of wave functions in Eq.(8) is conserved,
the small factor $D(A)$ should be compensated by the
contribution of quark configurations expanded over the large
volume $V$.  Particles with small momenta $p\sim1/V$ represent
just such configurations. So the probability of no pair creation
in $D(A)\to0$ fields is suppressed by the factor $\sim1/V$ to be
compared with the probability of configurations with quark pairs.

In order to make explicit the role of quarks with small momenta
for A and B fields, let us consider the number $n_R(p,t)$ of
quarks produced by such external field in the fermion vacuum:
$\Psi_0(t)=\exp(-iH_{ext}(t))|0>$. The simplest way to calculate
$n_R(p,t)$ is to use the bosonization procedure. For the fields
of the B-type these calculations have been performed in detail
in Sec.3 of Ref.[9]. To find $\Psi_0(t)$, one must solve the
Schr\"odinger equation in an arbitrary $A_0$ external field
taken in the Coulomb gauge. For this aim we express $\Psi(x)$,
$\Psi^+(x)$ operators by means of bosonization forms [9] and
calculate $n_R(p,t)$ as follows:
\begin{eqnarray}
 n_R(p,t)=\langle\Psi_0(t)|a^+_R(p)a_R(p)|\Psi_0(t)\rangle&=&\int
dxdy\,e^{ip(x-y)} \nonumber \\
\langle\Psi_0(t)|\Psi_R^+(x)\Psi_R(y)|\Psi_0(t)\rangle &=& \int
\frac{dxdy}{2\pi i}\frac{e^{ip(x-y)}}{y-x-i0} \nonumber \\
&\times& \exp\{2\pi i[\alpha(x,t)-\alpha(y,t)]\}.
\end{eqnarray}
We use the normalization condition to be
$\{a^+(p),a(p')\}=V\delta_{p,p'}$. In this case we have
\begin{eqnarray}
\alpha(x,t)&=&-\frac g{2\pi}\int\limits^t_0 dt_1\int dx_1
\Theta(t-x
-t_1+x_1)E(x_1,t_1)=\int\frac{dk}{2\pi}\frac{e^{-ik(t-x)}}k
\Phi_R(k,t) \nonumber \\
\Phi_R(k,t)&=&g/2\pi\int dt_1\int dx_1 e^{ik(t_1-x_1)}E(x_1,t_1).
\end{eqnarray}
For small $k$ we have for A and B types respectively:
\begin{equation}
\Phi_R(k,t)=i{\rm sign }k\,\beta(t), \qquad \Phi_R(k,t)
=-Q_{top}(t).
\end{equation}
Substituting Eqs. (14) and (15) to Eq.(13) and taking into
account the relation \begin{equation} \frac1{2\pi(y-x-i0)}=\int
\frac{dk}{2\pi}\,e^{-ik(y-x)}\,, \end{equation}  we obtain the
following formulae:
\begin{equation}
n_R(p,t)=\frac{sh^2\pi\beta(t)}p\quad (A), \qquad n_R(p,t) =
\frac{\sin^2\pi Q_{top}(t)}p\quad (B).
\end{equation}
Thus, for non-integer $Q_{top}$ both A and B type fields produce
an infinite number of $q\bar q$-\-pairs with small momenta
$p\to0$.  In the case of integer $Q_{top}$, Eq.(17) is not
defined when $p\to0.$ To obtain the correct expression for
$n_R(p,t)$ in the above case, let us consider the system in a
finite space volume $V$. Then, Eq.(13) must be changed as
\begin{eqnarray}
n_R(p_n,t)&=&\int\limits^{V/2}_{-V/2}dxdye^{ip_n(x-y)}
G^{(0)}_R(x-y)
\exp\left[\frac{2\pi}V\sum_{k_n>0}\frac{\Phi_R(k_n,t)}{k_n}
\right.\nonumber\\
&\times & \left.\frac{}{}\left(e^{-ik_n(t-x)}-e^{ik_n(t-x)} +
e^{ik_n(t-y)}  - e^{-ik_n(t-y)} \right)\right],
\end{eqnarray}
$p_n,k_n=2\pi n/V$ to be an integer. Furthermore, in the
considered case
the free Green function $G_0$ has the following form:
\begin{equation}
G^{(0)}_R(x-y)=\frac1V\sum_{p_n>0}e^{ip_n(x-y)}=
\frac{\exp(2\pi i(x-y)/V)}{V[1-\exp(2\pi i(x-y)/V)]}\ .
\end{equation}
Due to the $\{a^+(p),a(p)\}=V$ relation,
to obtain the number of particles $N_R(p_n,t)$
with the given momentum $p_n$ we must divide $n_R(p_n,t)$ by
volume $V$.  Finally, we substitute instead of $\Phi_R(k_n,t)$
in Eq.(18) its value at $k_n=0$ equal to $Q_{top}(t)$ and
obtain:  \begin{eqnarray}
N_R(p_n,t)&=&\frac1{V^2}\int\limits^{V/2}_{-V/2}dxdy
\frac{e^{ip_n(x-y)}e^{2i(x-y)/V}}{1-\exp 2i(x-y)/V}  \\
&\times&\exp\left[-\frac{2\pi Q_{top}}V\sum_{k_n>0} \frac1{k_n}
\left(e^{-ik_n(t-x)}+e^{ik_n(t-y)}-{\rm h.c.}\right)\right].
\nonumber
\end{eqnarray}
Summing up the series in the exponential factor we find
\begin{equation}
N_R(p_n,t)=\int\limits^{V/2}_{-V/2}\frac{dxdy}{V^2} e^{ip_n(x-y)}
\frac{\exp[2\pi i/V(x-y)]\exp[2\pi iQ_{top}(x-y)/V}{1
-\exp[2i(x-y)/V]}.
\end{equation}
Hence, for $Q_{top}<0$ we get:
\begin{equation}
N_R(p_n,t)=\left\{ \begin{array}{ll}
0 & p_n \ge| Q_{top}| \; 2\pi/V \\
1 & p_n=0, \; 2\pi/V,\ldots, 2\pi(|Q_{top}|-1)/V\ . \end{array}
\right.
\end{equation}
We see that B-type field with an integer $Q_{top}<0$ produces
exactly $|Q_{top}|$ of $R$-quarks with vanishing momenta 0,
$2\pi/V,\ldots,2\pi/V,$ $(|Q_{top}|-1)2\pi/V$. Simultaneously,
the same field produces the equal number $|Q_{top}|$ of $L$
antiquarks with identical momenta. For $Q_{top}>0$ quarks and
antiquarks change places. Namely, in this case, $Q_{top}$ of the
$L$ quarks and $Q_{top}$ antiquarks are produced.

The chiral properties of fermion systems in an external field
were investigated in Ref.[9] just for the case of B-fields. The
total chirality is changed according to the ABJ-\-anomaly [14]
as
\begin{equation} K(t)-K(0)=2Q_{top}(t).
\end{equation}
Here $K(t)=\int dxj^5_0(t,x)$ is the total chirality at time $t$,
$j^5(x)$ being the axial current density.

The explicit calculations [9] for $j^5_0(x,t)$ show that in the
case of a finite volume system,  there appear two different
components in the density of chirality when a B-field acts. The
first  component stays finite at $V\to\infty$. The second
component contains  terms $\sim1/V$ vanishing with $V\to\infty$
but giving a nonzero result  when they are integrated over $V$.
These terms appear due to the discussed above phenomenon:
particles with small momenta $p\sim1/V$ are necessarily produced
by B-fields. So the corresponding part of the total chirality
$-2(Q_{top})$ becomes delocalized over the volume $V$. The
localized part being formed by contributions of
quarks-\-antiquarks with finite momenta, is not conserved and
has to obey just Eq.(23). But the total chirality (including
the delocalized part) is conserved in the volume $V$.\footnote{
It means that the equation for the divergence of axial current
(ABJ-\-anomaly) changes for the case of finite volume as follows
(Ref.[9]):  $ ip_n\rho(p_n,t)+j(p_n,t)=-g/2\pi E(p_n,t),\;
n\neq0,\; j(p_n\neq0,t)=\dot K=0. $ with $(p_n=2\pi n/V)$. So
the total chirality being the particle number $(N_R-\bar
N_R)-(N_L-N_I)$, is conserved for the fermion system in the
external fields.} For $V\to\infty$ the described phenomenon
represents the physical interpretation of the ABJ-\-anomaly for
QED$_2$ in the Coulomb gauge.

Moreover, this effect provides both the ABJ-anomaly and
QED$_2$ chiral condensates. The matter is the following. When
$Q_{top}$ is an integer number it is $(Q_{top}|$ of $q_R\bar
q_L(q_L\bar q_R)$ pairs to be delocalized with $p\sim1/V$. In
the case with an external field when $Q_{top}$ is non-\-integer,
in  a coherent state with the distribution of Eq.(17) a
non-\-integer number of quarks is delocalized. On the other
hand, QED$_2$ models have chiral vacuum condensates only when
$|Q_{top}|$ is integer for essential self-\-consistent fields
(see next Sec.). This situation exists in the Schwinger model
where $|Q_{top}|=1$ and the condensate is formed mainly by
$q_R\bar q_L(q_L\bar q_R)$ complexes. For the QED$_2$ model with
many electron flavours [4,6] there are no condensates without
any symmetry between flavours. But only in the case of symmetry
the value $Q^{(f)}_{top}$ becomes integer on important
self-\-consistent fields for every flavour $f$. The vacuum
complexes exactly repeat the flavour symmetry in their structure
(for example, a symmetrical condensate $\Pi^{N_1}_{f=1}q_R^f\bar
q^f_L$ or $\Pi^N_{f=1}\bar q_R^f q^f_L$ for the $SU(N)$-\-case).
If there is no any flavour symmetry, there are no condensates.
But chiral complexes with $\ln V$ particles corresponding to
Eq.(17) $(\int dp/p\sim\ln V)$ are present here in vacuum
states. We have quantitatively investigated these questions in
Refs.[7,9].

Some properties of A and B fields are very similar  to instanton
fields in QCD [15]:
\begin{description}
\item[i)]  Dirac determinant $D(A)$ vanishes on the instanton
field:
\item[ii)]  Euclidean winding number $Q_{top}\neq0$;
\item[iii)] an instanton field also necessarily produces massless
quark pairs [16];
\item[iv)]  total chirality is changed by the instanton field
according to Eq.(23).
\end{description}
But in the case  of instantons $D(A)$ vanishes due to the zero
mode of the Dirac equation. Therefore, the Green function of a
massless quark become s infinite. That changes the quark
chirality. The quark functions stay finite in $B$-fields of
QED$_2$. For this reason, the chirality, in fact, appears to be
the conserved number.  EM-\-fields can produce in QED$_2$ only
$q_R\bar q_R$ and $q_L\bar q_L$-pairs, but $2|Q_{top}|$ of these
particles appear immediately in the unobserved vacuum
condensate.

The  essential fields were studied for QED$_2$ also in Ref.[12].
The consideration is based on the Euclidean functional integral
method.  Our consideration performed in the Minkowsky space
and the Coulomb gauge to be used indicates an importance of EM
field configurations different from those discussed in
[17].

\section{Physical role of the $D(A)\to0$ fields
in Schwinger model}

We show in the present Section that EM-fields where $D(A)\to0$
are responsible for
\begin{description}
\item[A)] both the
absence of charged particles and for screening of charges;
\item[B)] hadron states  to have a definite chirality $(K=0)$
in spite of the existence of the chiral condensate.
\end{description}

As it was stated in Sec.2, Gauss integrals for matrix elements
of the $S(T)$-\-operator are evaluated by a substitution of a
saddle-\-point field $\bar A$. This field depends on positions
of particles in initial and final quark configurations. Those
are the configurations determining investigated matrix elements.
The formula for saddle-\-point field is obtained in Ref.[10]
(Eq.(59) of Ref.[10]).  One has to consider only small momenta
$p\ll m$ as the results of Sec.2 witness. In the case of total
charge $Q=0$ we have the following equation in the
$p\rightarrow0$ limit:  \begin{eqnarray} A_0(p,t)&=&i\frac gm
\frac{R_i(p)+R_f(p)}|p| \frac{\cos m(T/2-t)}{\sin m\cdot T/2}
\nonumber \\ &-& i\frac gm\frac{R_f(p)-R_i(p)}|p| \frac{\sin
m(T/2-t)}{\cos m\cdot T/2}\ .  \end{eqnarray} $R_i(p)$ and
$R_f(p)$ are the sources depending on initial and final quark
configurations:
\begin{eqnarray}
R_i(p)&=
&\sum_k\Theta(-p)\left(e^{-ip\tilde x_k}-e^{-ip\tilde x'_k}
\right)+\Theta(+p)\left(e^{-ip\tilde y_k}-
e^{-ip\tilde y'_k}\right)\
,\nonumber \\
{}{}\\
R_f(p)&=&\sum_k\Theta(-p)\left(e^{-ipx_k}-e^{-ipx'_k}\right)+
\Theta
(p)\left(e^{-ipy_k}-e^{-ipy'_k}\right)\ ,\nonumber
\end{eqnarray}
where $\tilde x,\tilde x'\,(\tilde y,\tilde y')$ are coordinates
of $R(L)$ quarks and antiquarks in an initial state;
$x,x'\,(y,y')$ are those in a final state. The second term in
Eq.(24) is inessential
both for Green functions and $D(A)$ because it vanishes being
integrated over $t$. So it can be omitted.

If charged particles exist, then charges at the large distance
$r$ (for example, $q$ and $\bar q$) inevitably would contribute
to the wave function of two charged particles. Such $S(T)$
matrix element will be independent on $r$ within the $r\ll T$
time interval , i.e. within the region
\begin{equation}
|x_0-\tilde x_0|\sim|x'_0-\tilde x'_0|\sim T\ll r, \qquad
|x_0-x'_0|\sim r\ .
\end{equation}
But the moving $(\tilde
x_0\to x_0)$ $R$-quark contributes to the matrix element with
the field strength:  \begin{equation} E^{(q)}(p,t)=\frac
gm\left[\Theta(p)\,e^{-ip\tilde x_0}-\Theta(-p)\,
e^{-ipx_0}\right]\frac{\cos m(T/2-t)}{\sin m\cdot T/2}
\end{equation} or in the coordinate space:
\begin{equation}
E^{(q)}(x,t)=\frac{ig}{2\pi m}\left[\frac1{x-\tilde x_0+i0} +
\frac1{x-x_0-i0}\right]\frac{\cos m(T/2-t)}{\sin m\cdot T/2}\ .
\end{equation}
This is just the field of the A-type. To obtain the field
strength for an $R$-\-antiquark we take $x_0\to x'_0$, $\tilde
x_0\to\tilde x'_0$ and change the general sign in Eqs. (27) and
(28).

Substituting Eq.(27) into Eqs. (7),(9) we obtain that the
$q_R\bar q_R$ configuration, contributing to $S(T)$ as
(Eqs.(3),(9)):
\begin{equation}
b_R(\tilde x'_0)a_R(\tilde
x_0)G^{(T)}_R\left( T,\tilde x'_0;0,\tilde x'_0\right)D(\bar
A^{(q+\bar q)})G^{(T)}_R(T,x_0;0,x'_0) a^+_R(x_0)b^+_R(x'_0)
\end{equation}
decreases at $r\to\infty$ as
$r^{-2}\left(D(A^{(q)})\sim r^2,\;G\sim r^{-2}\right)$. Let us
note that $D(\bar A^{(q)})$ does not vanish but increases with
$r\to\infty$. This is a direct consequence that $E^{(q)}$ is a
purely imaginary value   at $(x_0-\tilde x_0)$ of Eq.(28).
However it is shown in the Appendix that the result $r^{-2}$ can
be obtained also by integrating over the following real fields
of the A-type:
\begin{equation}
E(p,t)=i\beta\left[\Theta(p)e^{-ip\tilde
x_0}-\Theta(-p)e^{-ipx_0} \right]\frac{\cos m(T/2-t)}{\sin
m\cdot T/2}\ ,
\end{equation}
where $\beta$ is a real parameter. The $r$ dependence in the
discussed amplitude is due only the integration over the $\beta$
parameter in (30). In this case we have that
$D(A)\sim1/r^{2\beta^2}$ and $G_{R,L}\sim1/r^{2i\beta}$, the
Jacobian being $\sim\sqrt{\ln r}$. Therefore, the integration
over fields, Eq.(30), gives the same result as Eq.(28):
\begin{equation}
\int_{0}^{\infty} d\beta\sqrt{\ln
r}\,\frac1{r^{2\beta^2}}\,\frac1{r^{4i\beta}}\sim\frac1{r^2}
\quad {\rm at}\quad r\rightarrow\infty.
\end{equation}

Thus, the screening mechanism is based on the fields of the
class A with $D(A)\to0$. It means that $S(T)$ contains only
configurations with $q$ and $\bar q$ near each other. Additional
particles should be present in configurations if some quark
situated far from some antiquark. Their role is to compensate
charges of distant $q\bar q$ particles. We have seen in the
previous section that the smallness of $D(A)$ can be compensated
by a contribution of quarks with small momenta $p\sim1/r$.  One
can demonstrate this phenomena for the considered case. For this
aim let us substitute the EM field, Eq.(27), into the factor
\begin{equation}
\int dx\,dx'
G^{(T)}_R(x,T,x',T)a^+_R(x)b_R^+(x')
\end{equation}
representing in $S(T)$ an additional $q_R\bar q_R$-pair in the
final state.  If the pair creation happens near the quark
$x_0\sim\tilde x_0$, we can neglect the influence of the
antiquark $(x'_0-\tilde x'_0)$ field: $x\gg|x-x_0|-|x'-\tilde
x_0|\sim|x_0-\tilde x_0|\sim T_0$. Using (7), we obtain the
Green function $G_R^T(x,T;x',T)$ for the field (27). In the
region $|x-x_0|\gg|x'-x_0|\sim T$ this Green function becomes:
\begin{equation}
G^{(T)}_R(T,x;T,x')=\frac1{2\pi
i}\exp\left[\int\limits^T_0 dt_1 \frac{\cos m(T.2-t_1)}{\sin
m\cdot T/2}\,\ln\frac1{T-t_1-x'+x_0-i0}\right]\ .
\end{equation}
In evaluating Eq.(33) we use the following
formula $$  \int G^{(0)}_R(x,T;x_1,t_1)dx_1dt_1A_0(x_1,t_1)=\int
\frac{dx_1dt_1}{2\pi i}\frac{A_0(x_1,t_1)}{T-x+x_1-t_1-i0} $$
\begin{equation}
=\; -\int\frac{dx_1dt_1}{2\pi i}\,\ln(T-x+x_1-t_1)E(x_1,t_1)\ .
\end{equation}

The Green function (33) does not depend on the quark coordinate
$x$. It depends only on the antiquark coordinate $x'$. Thus the
additional $q_R\bar q_R$-\-pair in the wave function (29)
represents a quark with the zero momentum $(\int
a^+_R(x)dx=a^+_R(p=0))$ and an antiquark with a finite momentum
determined by the field strength $E^{(q)}(x,t)$. The quark $x$ is
situated for from the original one $(x_0\sim\tilde x_0)$, while
the antiquark $x'$ is near $x_0$, and, hence, its charge
participates in the screening. The same effect happens with
$L$-Green functions.

Thus A-fields prohibit the existence of charged configurations
and inevitably provide the last ones with local screening
charges.

Now we proceed with the B-type fields. They maintain the
chirality conservation of any quark configurations. In order to
prove the statement let us consider an appearance of an
additional pair $a^+_R(x_0)b^+_L(y'_0)$ in the final state, i.e.
the changing of chirality by two. Eq.(24) shows that such a pair
produces the B-type field:
\begin{equation}
E^{(c)}(p,t)=-\frac
gm\left[\Theta(-p)e^{-ipx_0}+\Theta(p)
e^{-ipy_0}\right]\frac{\cos m(T/2-t)}{\sin m(T/2)}\ ,
\end{equation}
$$ Q^{(c)}_{top}(T)=-\frac g{2\pi}\int^T_0 dt\,E^{(c)}(0,t)=1 \ .
\eqno{(35')} $$
The B-field nullifies $D(A)$ and prohibits the suggested
process. It generates additional $qq$-\-pairs. Indeed, let us
consider the factor representing two such pairs:
$a_R^+(x)b_R^+(x')$ and $a^+_L(y)b^+_L(y')$. The wave function
acquires the factor
\begin{equation}
\int
dx\,dx'dy\,dy'a^+_R(x)G^{(T)}_R(x,T;x',T)b^+_R(x')a^+_L(y)
G^{(T)}_L (y,T;y',T)b^+_L(y')\ .
\end{equation}
For the fields (35), the Green functions $G_R$ and $G_L$ do not
depend on $x$ and on $y'$ respectively  Thus $E^{(c)}(p,t)$
produces a chiral pair $\bar q_R q_L$ near the original chiral
pair and $q_R\bar q_L$-\-particles with small momentum to be far
from the region $E^{(c)}\neq0$. The first chiral pair
compensates the change of the local chirality carried by an
the original one.  The existence of the chiral
condensate seems to be very natural when the creation of chiral
pairs with vanishing total momenta appear.
As it was explained in Sec.2, condensates really arises in
QED$_2$ model when $Q_{top}$ is an integer number for all
essential B-fields. Such conditions exist in the Schwinger model
(see Eq.(35')), and in the many electron model [4,6] with some
symmetry between flavours.

Let us note that B-type fields prohibit the change of chirality,
but states with the definite chirality can exist. For example,
the configuration with a $q_R\bar q_L$ pair both in the initial
and in the final states produces the field:
\begin{equation}
E(x,t) =
\frac{ig}{2\pi m}\frac{\cos m(T/2-t)}{\sin m\cdot T/2}\left[
\frac1{x-\tilde x_0+i0}-\frac1{x-\tilde y'_0-i0}+
\frac1{x-x_0-i0}-\frac1{x-y'_0+i0}\right].
\end{equation}
For this field $D(A)$ is non-zero and additional pairs need not
be produced. These quark configurations form the chiral
condensate in the Schwinger model [7].

To summarize, we see that localized states only with the definite
chirality can be stationary in QED$_2$. The chirality exchange
between them and the vacuum condensate is impossible. As we have
already noted that is the property which makes possible the
confinement of fermion numbers in QED$_2$.

\section{Space-time picture  of charge
screening and of the hadronization in physical processes}

The production of  a chiral quark pair  $q_R\bar q_L(\bar
q_R,q_L)$ by an external source is an analogy of the $e^+e^-$
hadron process in QED$_2$. The process had been studied in
Refs.[5,8,9]. The formulae for EM charge densities
$\rho_{R,L}(x,t)$ for $R$- and $L$-\-particles were obtained
there to be
\begin{equation}
\rho_{R,L}(x,t)=\pm\left[\delta(t\mp x)-\frac m2\frac{t\pm
x}{\sqrt{t^2-x^2}}J_1\left(m\sqrt{t^2-x^2}\right)\Theta(t^2-x^2)
\right] \,,
\end{equation}
$J_1$ being the Bessel function. One can observe in Eq.(38) the
leading and the screening charges. The both are situated near
the light cone.  The screening charge is accumulated in the
region $\Delta x\sim1/m^2t$ near the leading particle. After the
time $t\sim p_{in}/m^2$ ($\Delta x\sim p_{in}^{-1}$ and $x\sim
m^{-1}$ in the restframe) the leading quark starts to interact
with the screening ones. The leading quark moves as a free
particle till times $t\le p_{in}/m^2$ [5,8].

We consider in this section quantitatively all three stages for
evolution in time of the parton process: a)~neutralization of
jets, b)~accumulation of partons, c)~transition into hadrons.
The physics is closely connected with the field generated by a
$q_R\bar q_L$ pair, i.e. with the B-type field.

Let us formulate the results.

1. $R$ and $L$ jets practically do not effect with each other
after the time $t_0\sim m^{-1}$, when particles to be already
generated neutralize total charges and chiralities of
each jet.

2. The neutralizing charges appear as a chiral pair $(\bar
q_Rq_L)$ with momenta $p\sim m$. This effect is the result of
B-field action.

3. Quarks of large momenta $p\gg m$ are accumulated in
almost neutral systems $(t>t_0)$. They exist only at the light
cone region. They softly turn into hadrons, which slow down
 $\sim p$, the hadronization time is $t\simeq p/m^2$.

4. Inside the light cone there are only quarks to have been
organized into hadrons. In the region between charges there are
no any string of quarks-\-antiquarks.

The screening of total jet charges ($q_R$ and $\bar q_L$) and
the transition into neutral hadrons require to change chirality
of the $q_R\bar q_L$-pair $(K=2)$. As we know (Sec.3) the field
of B-type must appear with such effect. Indeed, calculating the
field strength $E(x,t)$ for the distribution of Eq.(38) we
obtain
\begin{equation}
E(x,t)=g\,J_0\left(\sqrt{t^2-x^2}\right)\Theta(t^2-x^2)\ .
\end{equation}
It is the field of the B-type since at $p\to0$
\begin{equation}
E(p,t)=2g\frac{\sin mt}m\ .
\end{equation}
The corresponding shift of chirality is equal to
\begin{equation}
K(t)-K(0)=2Q_{top}(t)=2(\cos mt-1)\ .
\end{equation}
The chirality changing (41) implies the creation of two kinds of
quark pairs by the B-field (see Sec. 2--3).

1) One additional $\bar q_Rq_L$ pair with the momentum $p\sim m$.
These particles  compensate total charges and chiralities
of $R$ and jets. This pair exists after $t_0\sim1/m.$

2) One $\bar q_Rq_L$ pair with the total momentum $p\sim1/t$. It
is expanded always over the whole causal region $(2t)$. This
pair appears simultaneously with the first one, as chirality of
local QED$_2$ systems can be changed only on the time of
fluctuations $(\sim1/m)$. Chirality oscillations mean a
fluctuative exchange between just this pair and the condensate.
The pair joins the condensate at $t\to\infty$.

The latter pair has no connection with the parton accumulation
and transitions into hadrons. The both processes generate
independent and neutral $(Q=0,$ $K=0$) compact $R,L$ packets.
The quarks of the first $\bar q_Rq_L$ pair participate in the
hadron production. They form hadrons with $p\sim m$ at the time
$t\sim t_0\sim1/m$.

To prove the above assertions, we investigate the time
dependence  of the number of quarks produced by the field,
Eq.(39), in the considered process. The strict calculations by
means of the bosonization method gives [9] the answer very
similar to Eq.(13).  The density of the right-\-handed quarks is
\begin{eqnarray}
n_R(p,t)& =
&\langle\Psi(t)\beta^+_R(p)a_R(p)|\Psi(t)> \;= \\
&=&\int\frac{dxdy}{2\pi i}\frac{e^{ip(x-y)}}{y-x-i0}f(x-y)\left[
e^{2\pi i(\alpha(x,t)-\alpha(y,t))} -1\right]\ .\nonumber
\end{eqnarray}
Here, $\alpha(x,t)$ can be expressed in terms of the field strength
$E(x,t)$, Eq.(39), by means of Eq.(14). The function
\begin{equation}
f(x-y)=\exp\left[-\int\frac{dk}{w_k}\left(\frac{w_k-k}k\right)^2
\sin^2\frac{k(x-y)}2\right], \quad w_k=\sqrt{k^2+m^2}
\end{equation}
determines the mean density of the quarks in the vacuum.

The integrand in Eq.(42) depends on $\xi=t-x$ and $\xi'=t-y$. For
$t\gg t_0\sim1/m$ and $p\le m$ only the difference $\xi-\xi'$ is
restricted by a rapid decrease of the function $f(x-y)=f(\xi)$.
The values of $\xi$ and $\xi'$ themselves are large $(-t)$. In
this case the integration in Eq.(14) goes over all possible
$x_1$ and $t_1$. Thus, we have
\begin{equation}
\alpha(x,t)=-\frac g{2\pi} \Theta(t-x)\int^t_0dt_1\int
gJ_0(m)(t^2_1-x^2_1) \Theta (t^2_1-x^2_1)dx_1\ .
\end{equation}
Applying
Eq.(16) and introducing the variables $\zeta=\xi-\xi'$ and
$\xi'$, we obtain
\begin{equation}
n_R(p,t)=\int^\infty_p \frac{dx}{2\pi}\int
d\zeta,\,e^{-ik\zeta}f(\zeta)F(\zeta),
\end{equation}
where
\begin{equation}
F(\zeta)=\zeta\exp[i\pi Q_{top}(t)\varepsilon(\zeta)]2i\sin2\pi
Q_{top}(t)\ ,
\end{equation}
and $Q_{top}(t)$ is the same as in Eq.(41). Finally,
substituting Eq.(46) in Eq.(45) we get
\begin{equation}
n_R(p,t)=A(p)\sin^2\pi Q_{top}(t)+B(p)\sin2\pi Q_{top}(t)\ ,
\end{equation} and
\begin{equation}
A(p)=\frac2\pi \int^\infty_0 d\zeta f(\zeta)\sin p\zeta
, \qquad B(p) =\frac1\pi\int^\infty_0 d\zeta f(\zeta)\cos p\zeta
\ .
\end{equation}
Equation (47) clearly demonstrates the
oscillating nature of the density of quarks produced with small
momenta. The first term in Eq.(47) describes production of
quarks by the field given by Eq.(39). For
$f(\zeta)=e^{-\delta\zeta}$ $(\delta=0)$ we return to Eq.(17).
The second term describes the influence of the field on vacuum
quarks. That is the effect which produces oscillations.

$R$ quarks of Eqs.(42)--(48) are bounded into chiral pairs. Since
in Eq.(41) there are no restrictions on $x$ and $y$ values
$(|x|,|y|)\le t)$, the pairs are expanded over the whole volume
$\sim2t$, i.e. the total momentum of the pair is $p\sim t^{-1}$.
The momenta of the quarks inside one pair are determined by the
integrals in Eq.(48) and are of the order of $p\sim m$. The
total number of bounded quarks is
\begin{equation}
\int
A(p)\frac{dp}{2\pi}\sim\int B(p)\frac{dp}{2\pi}\sim1.
\end{equation}
Therefore we have dealt with one pair. The value of $Q_{top}$
oscillations also confirms this fact.

Let us now consider particles with momenta $p\gg m$. They do not
participate in transitions into vacuum, but just these particles
form hadrons. The last ones with momentum $P$ are formed during
the time
\begin{equation}
t\sim \frac P{m^2}\ .
\end{equation}
The size of the quark packet, Eq.(38), becomes then equal to the
size of such hadrons $\sim1/P$. After the time $\sim P/m^2$ all
quarks with momenta $p\le P$ have to be bounded. Equation (42)
gives the opportunity to verify this qualitative picture. In
some aspects we repeat Ref.[9].

For particles with $p\gg m$ the value of $|x-y|$ becomes
$\sim1/p$, and $f(\zeta)\sim f(0)=1$. These quarks do not
interfere with vacuum quarks. We apply again Eq.(16) and rewrite
Eq.(42) in the form ($\zeta=(t-x)$):
\begin{equation}
n_R(p,t)=\int^\infty_p
\frac{dx}{2\pi}\left|\int^{+\infty}_{-\infty} e^{-ik\zeta-2\pi
i\alpha(\zeta,t)}d\zeta\right|^2\ .
\end{equation}
Since jets become neutral at $t\sim1/m$ we have $n_R=\bar
n_R=n_L =\bar n_L$.

In Eq.(14) the essential integration region is near to the light
cone:
\begin{equation} 0<\zeta_1=t_1-x_1<\zeta=t-x\sim\frac1p\ .
\end{equation}

Therefore, quarks with momenta $p$ are produced only in the
region $1/p$ near the leading particles. Inside the light cone
there are quarks with $p<m^2t$. They have already formed
hadrons.

According to Eq.(52) the $t^2_1-x^2_1$ interval is approximately
$t^2_1-x^2_1=2t_1\zeta_1$, and $\alpha$ is
\begin{equation}
\alpha(\zeta,t)=-\frac{m^2}2\int
dt_1\int^\zeta_0d\zeta_1J\left(m\sqrt{2t\zeta_1}\right)=e(\zeta)
\left[J_0\left(m\sqrt{2t\zeta}\right)-1\right]\ ,
\end{equation}
since
$\Theta(t^2_1-x^2_1)\Theta(\zeta-t_1+x_1)=\Theta(\zeta_1)
\Theta(\zeta-\zeta_1)$. Eq.(53) shows that jets are independent
from each other for $(t>1/m)$ because only the field of the
$R$-jet produce $R$-quarks.  In the time interval $m^{-1}\ll
t\ll p/m^2$ we can expand the Bessel function in Eq.(53) and get
\begin{equation}
n_R(p,t)=\frac\pi6\left(\frac{m^2t}p\right)^2\frac1p,
\qquad m^{-1}<t<\frac p{m^2}\ .
\end{equation}
At times $t>p/m^2$ the
argument of the Bessel function  becomes large and we expand the
exponential factor in Eq.(51) in powers of $\alpha(\zeta,t)$.
The integrations over $k$ and $\zeta$ give the distribution
to be independent of time as follows
\begin{equation}
n_R(p,t)=\frac{2\pi}p\
, \qquad t\gg\frac p{m^2}\ .
\end{equation}
At times $t\sim p/m^2$ the
distribution of Eq.(54) becomes the order of unity. Quarks with
the momentum $p$ begin to turn into hadrons.  Eq.(55) represents
the distribution of quarks inside hadrons.

Indeed, we have in the parton model
\begin{equation}
n_R(p,t)=\int N(P,t)n_R(P,p)\,\frac{dP}{2\pi}\ .
\end{equation}
Here $N(P,t)$ is the number of hadrons with the momentum $P$. In
the process considered this quantity is equal to [5,9]:
\begin{equation}
N(P,t)=\frac{\langle\Psi(t)|A^+(p)A(P)|\Psi(t)\rangle}{\langle
\Psi(t)\mid\Psi(t)\rangle}=\frac{2\pi}{\omega_p}\ .
\end{equation}
Here, $A^{\pm}(p)$ are hadron operators. The distribution of
quarks inside a hadron with the momentum $P$ is equal to:
\begin{equation}
n_R(P,p)=\langle\Psi_h(P)|a^+_R(p)a_R(p)|\Psi_h(p)\rangle
=\frac{2\pi}P\ .
\end{equation}
We obtain Eq.(58) by means of the parton wave function for a
QED$_2$ hadron. We had found in Refs. [9,10] that
\begin{eqnarray}
\Psi_h(P)=
\sqrt{\frac{2\pi}P}\int\frac{dk}{2\pi}a^+_n(P-k)b^+_R(k)
|\Omega_0>\ ,\qquad P\gg m \nonumber\\
a_{R,L}(k)|\Omega_0> =b_{R,L}(k)|\Omega_0>=0\ , \qquad {\rm
for}\quad k\gg m\ .
\end{eqnarray}
Substituting Eqs. (57)--(58)
in Eq.(56) we obtain the distribution of Eq.(55).

The leading particles lose their energies and turn into hadrons
after the time $t\sim P_{in}/m^2$. Hence, the most rapid hadrons
are produced in the last turn (it is similar to Refs.[5,17]).

\section{Conclusion}

In the present paper it was demonstrated that confinement of the
charged states in QED$_2$ is provided by the field configurations
which nullify the Dirac determinant while Green functions stay
finite in these fields. It is possible that this confinement
mechanism can exist not only in two dimensions. But at present
it is unknown whether the fields with similar properties play
any role in QCD, in four dimensions. But if the $D(A)\sim0$
fields are important also in QCD, the mechanism based on such
fields would be very attractive because it naturally provides
theory with many phenomenological inviting properties.

In fact, the colour screening in QCD could appear because
quark-antiquark create a coloured field providing for
the $D(A)$ determinant to go to zero when the distance
between them increases.  This field produces (with probability
of one) soft quarks  which screen the initial colour.  These
quarks create the baryon number current which could,
generally, screen also the triality of the sources.

Certainly, the screening is much more complicated for QCD
because of its non-\-Abelian nature (also we must screen the
gluon colour) and probably needs more time than in QED$_2$.
However in the described  scenario the momenta of leading quarks
are much more than momenta of the screening ones. So it seems
that the time of screening would be less than the time of
hadronization. As a consequence hadrons are produced after the
time when the screening of colour is over and jets are
independent and colourless. Thus the fast hadrons (which are
produced the last) would be created quite independently in
different jets.  Light quarks
would play a fundamental role in this confinement mechanism.

\appendix\section{}

The explanation concerning the imaginary saddle field in
Eqs.(28),(29) consists in the explicit evaluation of the
functional integral. We have for the matrix element considered:
\begin{equation}
S_0\equiv S^{q\bar q}_{q\bar q}(T)=\int DA(x)\exp[iS(A)]G^{(T)}_R
\left(T,x'_0;0\tilde x_0'\right)D(A)G^{(T)}_R(T,x_0;0,\tilde
x_0)\ .
\end{equation}
We are interested only in components of $A$ with
very small $p$, these ones are important for the dependence on
the distance $r$ between $q$ and $\bar q$ when $r\to\infty$.
Then we have \begin{equation}
S(A)=\int\limits^{P_0}_0\frac{dp}2\left\{\frac{im^2|p|}
4 a(p)a(-p)
+ig[R_1(p)+iR_2(p)]a(-p)\right\}+S(p\gg p_0)\ ,
\end{equation}
where
$$ p_0\ll m, \qquad a(p)=\int^T_0 A_0(p,t)dt\ .\eqno{(A.2')} $$
We neglect in Eq.(A.2) the kinetic energy  of the small
components:  $\int^{P_0}_0 p^2A^2_0(p,t)dp/2\pi$. We take the
sources from Eq.(25) and rewrite down them for the $q\bar
q$-\-case in the following way
\begin{equation}
R_i(p)+R_f(p)=R_1(p)+R_2(p)\ .
\end{equation}
$R_1(p)=R_1^*(-p)$ and $R_2(p)=-R^*_2(-p)$ are real and imaginary
parts of the sources.

The integral can be easily transformed to
\begin{eqnarray}
S_0=Z\lambda\int d\beta e^{ig\lambda\beta}\int
Da(r)\delta\left(\int R_1(p)a(p)\frac{dp}{2\pi}-
\beta\lambda\right)
\nonumber \\
\exp\left\{i\int\frac{dp}{2\pi}\left[\frac{im^2|p|}4 a(p)a(-p)
+igR_2(p)a(-p)\right]\right\} G_RG_RD\ .
\end{eqnarray}
Here $Z$ includes all contributions independent on $r$, and
\begin{equation}
\lambda=
\int\frac{R_1(p)R_1(-p)}{|p|}\,\frac{dp}{2\pi}\sim \ln r\ .
\end{equation}
Let us change the integration variables according to
\begin{equation}
a(p)=\beta\frac{R_1(p)}p+\tilde a(p)=a_0(\beta,p)+\tilde a(p)\ .
\end{equation}
We obtain
\begin{eqnarray}
S_0= Z\lambda\int d\beta\,D(a_0)G_R(a_0)G_R(a_0)\cdot I \\
I=\int D\tilde a(p)\delta\left(\int R_1\tilde
a\frac{dp}{2\pi}\right)
\exp\left[-\int\frac{dp}{2\pi}\frac{m^2|p|}4\tilde a(p)
\tilde a(-p)
\right]\ .\nonumber
\end{eqnarray}
The dependence on $R_2$ is inessential for $r\to\infty$. Thus the
integration in Eq.(47) goes over real fields and these fields
are of the A class. Practically we finish the proof of our
assertion in the text. The further evaluation of $I$ is
straightforward.
\begin{eqnarray}
I&=&\int D\tilde a\int\frac{d\zeta}2\exp\left[-\int\frac{dp}
{2\pi}
\frac{m^2|p|}4\tilde a(p)\tilde a(-p)+i\zeta R_1(p)\tilde a(-p)
\right] \nonumber \\
&=& \int^{+\infty}_{-\infty}\frac{d\zeta}{2\pi}\exp\left[
-\frac{\zeta^2}{m^2}\,\lambda\right]\sim \frac1{\sqrt{\ln r}}\ .
\end{eqnarray}
Substituting all factors in Eq.(A.7) we get the result of the
text.  Obviously, the saddle point for integration over $\beta$
is imaginary: $\beta=i$, and we return to the field of Eq.(22).

\end{document}